\documentclass[reprint,superscriptaddress,amsmath,amssymb,prl]{revtex4-2}
\usepackage{color}
\usepackage{amsthm,amsmath,amssymb}
\usepackage{mathrsfs}
\usepackage{dsfont}
\usepackage{braket}
\usepackage{graphicx}
\usepackage{dcolumn}
\usepackage{bm}
\usepackage{mhchem}
\usepackage{multirow}
\usepackage{booktabs}
\usepackage{diagbox}
\usepackage{makecell}
\usepackage{stackengine}
\usepackage[colorlinks=true,linkcolor=blue,
          citecolor=blue,urlcolor=blue,]{hyperref}

\begin{document}
\title{Observation of Quantized Klein Tunneling in a Dielectric Resonator Chain}

\author{Rui-Jie Zhang}
\thanks{These authors contributed equally to this work.}
\affiliation{Lanzhou Center for Theoretical Physics, Key Laboratory of Theoretical Physics of Gansu Province, Key Laboratory of Quantum Theory and Applications of MoE, and School of Physical Science and Technology, Lanzhou University, Lanzhou, Gansu 730000, China}

\author{Xiao-Zhen Peng}
\thanks{These authors contributed equally to this work.}
\affiliation{Lanzhou Center for Theoretical Physics, Key Laboratory of Theoretical Physics of Gansu Province, Key Laboratory of Quantum Theory and Applications of MoE, and School of Physical Science and Technology, Lanzhou University, Lanzhou, Gansu 730000, China}

\author{Ri-Zhen Yang}
\affiliation{Lanzhou Center for Theoretical Physics, Key Laboratory of Theoretical Physics of Gansu Province, Key Laboratory of Quantum Theory and Applications of MoE, and School of Physical Science and Technology, Lanzhou University, Lanzhou, Gansu 730000, China}

\author{Rui-Hua Ni}
\affiliation{Lanzhou Center for Theoretical Physics, Key Laboratory of Theoretical Physics of Gansu Province, Key Laboratory of Quantum Theory and Applications of MoE, and School of Physical Science and Technology, Lanzhou University, Lanzhou, Gansu 730000, China}

\author{Yong-Yin Hu}
\affiliation{Lanzhou Center for Theoretical Physics, Key Laboratory of Theoretical Physics of Gansu Province, Key Laboratory of Quantum Theory and Applications of MoE, and School of Physical Science and Technology, Lanzhou University, Lanzhou, Gansu 730000, China}

\author{Hong-Ya Xu}
\email{Corresponding author: xuhongya@lzu.edu.cn}
\affiliation{Lanzhou Center for Theoretical Physics, Key Laboratory of Theoretical Physics of Gansu Province, Key Laboratory of Quantum Theory and Applications of MoE, and School of Physical Science and Technology, Lanzhou University, Lanzhou, Gansu 730000, China}

\author{Liang Huang}
\email{Corresponding author: huangl@lzu.edu.cn}
\affiliation{Lanzhou Center for Theoretical Physics, Key Laboratory of Theoretical Physics of Gansu Province, Key Laboratory of Quantum Theory and Applications of MoE, and School of Physical Science and Technology, Lanzhou University, Lanzhou, Gansu 730000, China}

\date{\today}

\begin{abstract}
We present the first experimental observation of quantized Klein tunneling in a bounded Dirac system, implemented by a dimer chain of dielectric microwave resonators. Both the unusual quantized levels and corresponding spinor states hybridized from distinct particle and hole wavefunctions are measured. 
All observations are in quantitative agreement with the hitherto-untested
prediction of Dirac equation. 
Our results make an important step to realize and understand the particle-hole physics of Klein tunneling in bounded Dirac systems, and also shed light on potential applications for manipulating particle-hole hybridized spinor waves.
\end{abstract}
\maketitle

{\it Introduction.}---In 1929, Klein found, in his famous gedanken experiment, a massive Dirac particle can propagate through an arbitrarily high step potential without decay---now known as Klein tunneling~\cite{Klein1929, Dombey1999}. Two years earlier, Hund worked out the nonrelativistic tunnel effect for a bounded system, and recognized its relevance for hybrid orbitals and molecular spectra \cite{Hund1927, Merzbacher2002}. An intriguing problem is then to address Klein tunneling in bounded relativistic quantum systems.

As one of the most striking and counterintuitive consequences of Dirac equation, Klein tunneling has implications in broad contexts ranging from exotic situations of super-heavy nuclei \cite{1985Quantum,Grib1994} and black hole evaporation \cite{Page_2005}, to anomalous transport~\cite{Katsnelson2006, Beenakker2008, Young2009, Stander2009, Yang2011, Brien2016, Elahi2024} and quasi-bound states~\cite{Shytov2008,Zhao2015,Christopher2016,Lee2016,Zheng2023, Jois2023} in Dirac materials. 
Recently, tremendous interest has been aroused by its experimental realizations and promising applications in analog systems, such as graphene \cite{Elahi2024}, topological insulator~\cite{Lee2019}, cold atoms \cite{Salger2011}, trapped ions~\cite{Gerritsma2011}, acoustic \cite{Jiang2020DirectOO,Gao2022} and photonic \cite{Dreisow_2012, Zhang2022,Gao2023} crystals, and non-Hermitian lattices~\cite{yu2024dirac}. 
Despite transport and quasi-bound state related investigations, a direct observation of quantized Klein tunneling in completely confined Dirac systems remains elusive.

\begin{figure}[t]
\includegraphics[width=\linewidth]{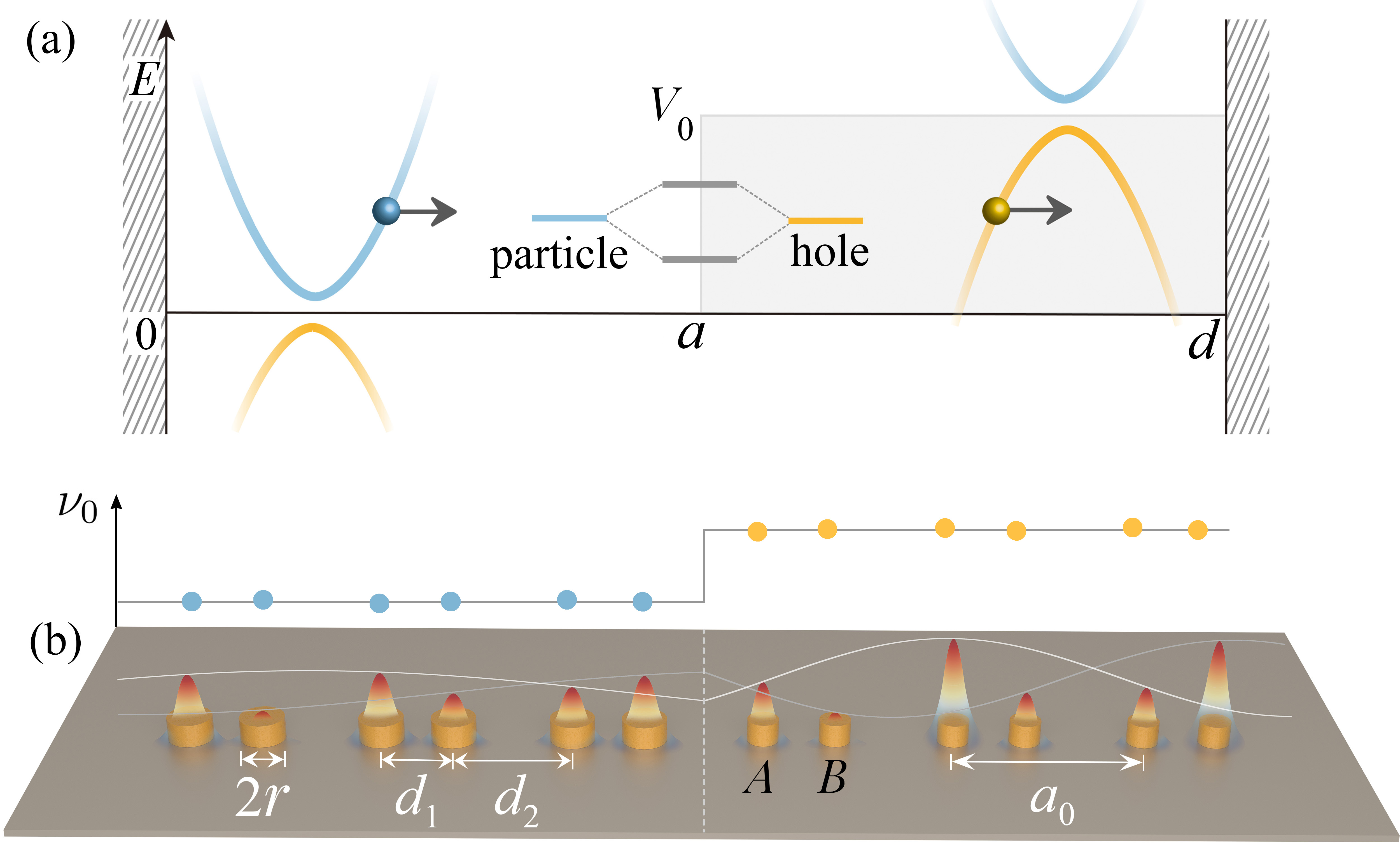}
\caption{\label{fig:01} (a) The bounded Dirac system. The step potential of height $V_0$ is to evoke Klein tunneling that hybridizes the particle states and the hole states to form the resulting bound states. (b) Schematics of the microwave analog. Cylindrical dielectric resonators with high permittivity $\varepsilon = 37$ and height 5 mm are placed as a dimer chain between two metallic plates of distance 13 mm (the top plate is not shown). The radius' change of the resonators from left ($r=4$ mm) to right ($r=3.93$ mm) is to create a sudden increase of the resonant frequency $\nu_0$ of uncoupled resonators, emulating the step potential. A quantized Klein tunneling state is superimposed onto the chain, where the curves are the outlines for $A$, $B$ sublattices.
}
\end{figure}

We tackle this issue in a bounded massive Dirac system with a step potential (Fig. \ref{fig:01}). The focus is on the Klein tunneling regime. Our model, incidentally, emulates a peculiar relativistic diatomic molecule \cite{2006Relativistic, dyall2007introduction, Peeters2019, Helin2020}, by viewing the left and right domains as artificial Dirac atoms with their own discrete spectrum. Klein tunneling is essential in forming the unusual discrete orbitals through coupling the particle-like states and the hole-like states (abbreviated as particle and hole hereafter). 

Here, using a microwave realization of the bounded massive Dirac system, we observe, for the first time, quantized Klein tunneling by direct accessing of the discrete levels and corresponding spinor states. These states are hybridized from distinct particle and hole wavefunctions with complementary wavevectors, exhibiting heterogeneous spinor waveforms. Note that the finite mass will degrade significantly the Klein tunneling due to an emerging gap if the applied step potential is not abrupt~\cite{Sauter1931, Gerritsma2011, tran2023klein}. This poses difficulties comparing to the massless case studied in most of existing experiments \cite{nakatsugawa2023massive}. Our scheme surmounts this challenge, which is the key to the excellent agreement between the data and the theory. 
 
These results are directly relevant to the understanding of the central role of Klein tunneling in shaping the quantized levels and states in bounded Dirac systems, and offer a versatile platform for potential applications by exploring relativistic particle-hole physics with the capability of tailoring hybridized spinor waveforms.

{\it Model.}---The Hamiltonian for a Dirac particle of mass $m$ confined in a one-dimensional box in $[0,d]$ is
\begin{equation}\label{eq:H}
  H=-{\mathrm i}c \hbar \sigma_x \partial_x + m c^2 \sigma_z + V(x)\mathds{1}_2,
\end{equation}
where $\sigma_x$ and $\sigma_z$ are Pauli matrices, $V(x)=0$ for $0\leq x<a$, and $V(x)=V_0$ for $a \leq x \leq d$ [Fig. \ref{fig:01}(a)]. 

\begin{figure*}[t]
    \centering
    \includegraphics[width=0.9\linewidth]{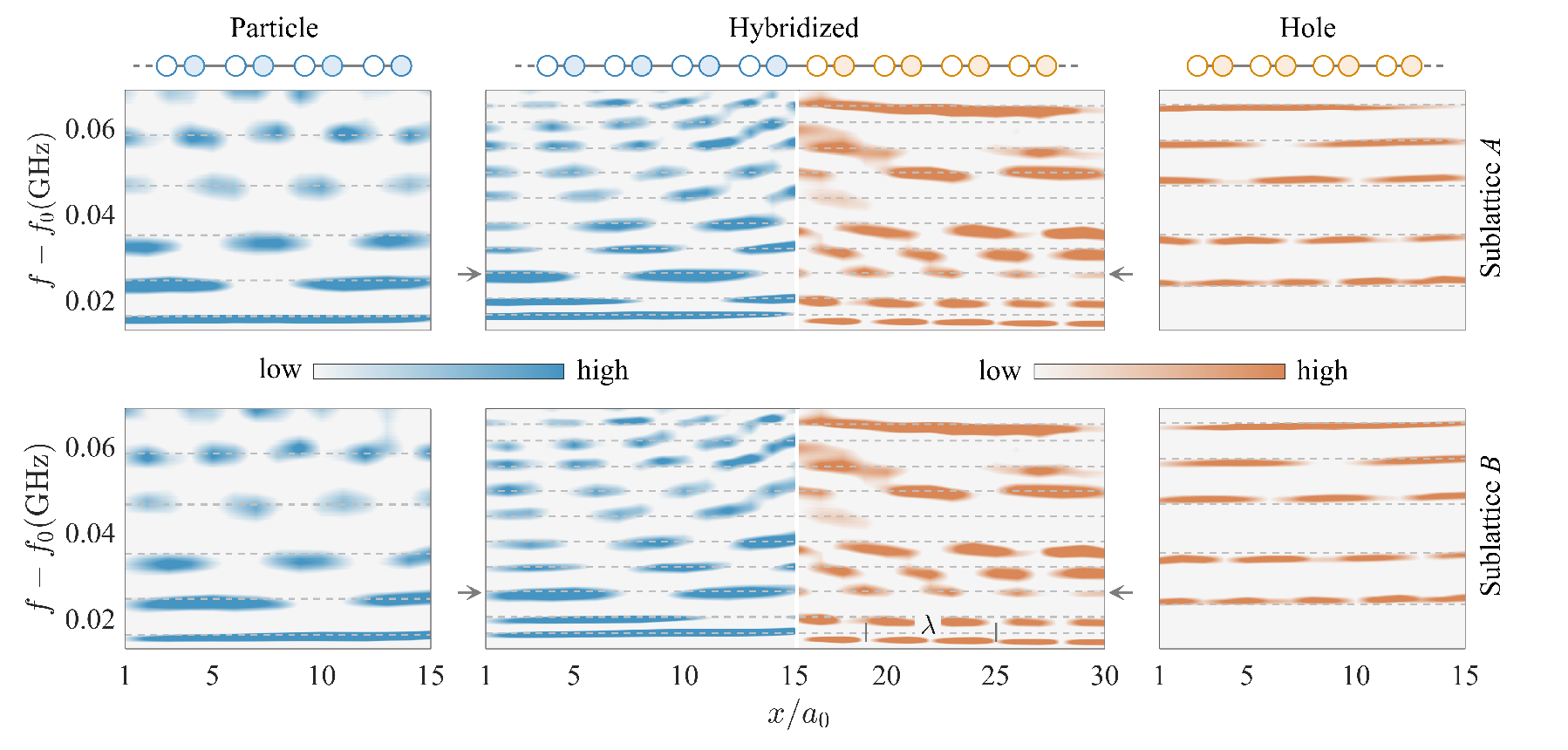}
     \caption{\label{fig:03} The LDOS for Experiment 1 in the regime of Klein tunneling. Upper (lower) parts are for $A$ ($B$) sublattices. Left and right columns are for particle states of the left 15 dimers alone and hole states of the right 15 dimers alone, respectively. Middle column is for the particle-hole hybridized states of the whole system, where the minimum wavelength $\lambda$ is about $7a_0$. The dashed horizontal lines represent the eigenlevels from the theory for each case with $f_n-f_0 =E_n$. The arrows indicate the third eigenstate that will be plotted in Fig. 5(a, b). Insets above the main panel show the schematics of the dimer chain of dielectric resonators for the three cases. } 
\end{figure*}

This system is simulated by the celebrated Su-Schrieffer-Heeger model \cite{SSH1979} using a chain of $N$ dimers of evanescently coupled dielectric resonators [Fig. \ref{fig:01}(b)] sandwiched between two metallic plates \cite{Franco2013, Poli2015, Cl0Wavefront}. 
With the intra- and inter-dimer couplings $v$ ($d_1=10$ mm) and $w$ ($d_2=10.5$ mm), the effective Hamiltonian around the band gap 
is \cite{Longhi2010, book2016}
\begin{equation}\label{eq:F}
  F= {\mathrm i} w a_0 \sigma_y \partial_x + (v-w)\sigma_x + f(x)\mathds{1}_2,
\end{equation}
where $a_0 = 20.5$ mm is the lattice constant, and $f(x)$ mimics the step potential. In practice, we choose a group of 30 cylindrical resonators with the most close bare resonant frequencies $\nu_0$ from each of the two sets, e.g., around $6.7014$ GHz for the $r=4$ mm resonators placing on the left side, and $6.7829$ GHz for the $r=3.93$ mm ones on the right side, where the standard deviation for each group is about 2.7 MHz. Note that $\nu_0$ determines mostly the frequency $f_0$ of the Dirac point for the dimer chain \cite{reff0}. Thus $f(x)$ has a sudden change $\Delta f$ within a spatial scale of $a_0/2$, forming an abrupt step as the minimum wavelength is about $7a_0$ in our experiments (c.f. Fig. \ref{fig:03}).

With equivalent parameters $mc^2=(v-w)$, $c \hbar = w a_0$, and $V_0 = \Delta f$, it is straightforward that the Hamiltonians are related by $F-f_0 = U H U^\dag$, where $U=(\sqrt{2}/2)[1, {\mathrm i}; 1, -{\mathrm i}]$ is a unitary matrix. The eigenvalues and spinor eigenwavefunctions of $F$ and $H$ are related by
\begin{equation}\label{Eq:correspondence}
f_n-f_0 = E_n \ \ \ {\rm and} \ \ \ \psi_n = U\Psi_n.
\end{equation}
The dimer chain has open boundaries, which is equivalent to assume additional resonators on both ends but with vanishing wavefunction \cite{book2016}. Thus $\psi^{(B)}_{j=0}=\psi^{(A)}_{j=N+1}=0$ ($\psi^{(2)}|_{x=0}= \psi^{(1)}|_{x=d} =0$) for the two sublattices (components). 
For the Dirac Hamiltonian $H$, since $\Psi=U^\dag \psi$, we have $-[\Psi^{(2)}/\Psi^{(1)}]|_0=[\Psi^{(2)}/\Psi^{(1)}]|_d={\mathrm i}$, i.e., the hard-wall infinite mass boundary with vanishing outgoing current \cite{berry1987neutrino}, a peculiar boundary condition to confine relativistic particles \cite{Chodos1974}. With $N_L$ ($N_R$) dimers on the left (right) side, the box length is then $d=(N+1/2) a_0$, and $a=(N_L+1/4) a_0$, where $N = N_L + N_R$. This establishes the correspondence between the experiment and the theory with no free parameters.

{\it Results.}---The resonators are placed on the copper plate by a high precision automatic robotic arm. The reflection spectra $S_{11}$ are measured using a vector network analyzer by positioning the loop antenna above each resonator. Following Ref. \cite{Bellec2013}, the local density of states (LDOS) can be derived from $S_{11}$, which can lead to the density of states (DOS) to determine the quantized eigenlevels, and also the wavefunction intensities $|\psi_n(x)|^2$ for separated resonances (Supplemental Material \cite{RefSM}).

\begin{figure}[t]
    \centering
    \includegraphics[width=\linewidth]{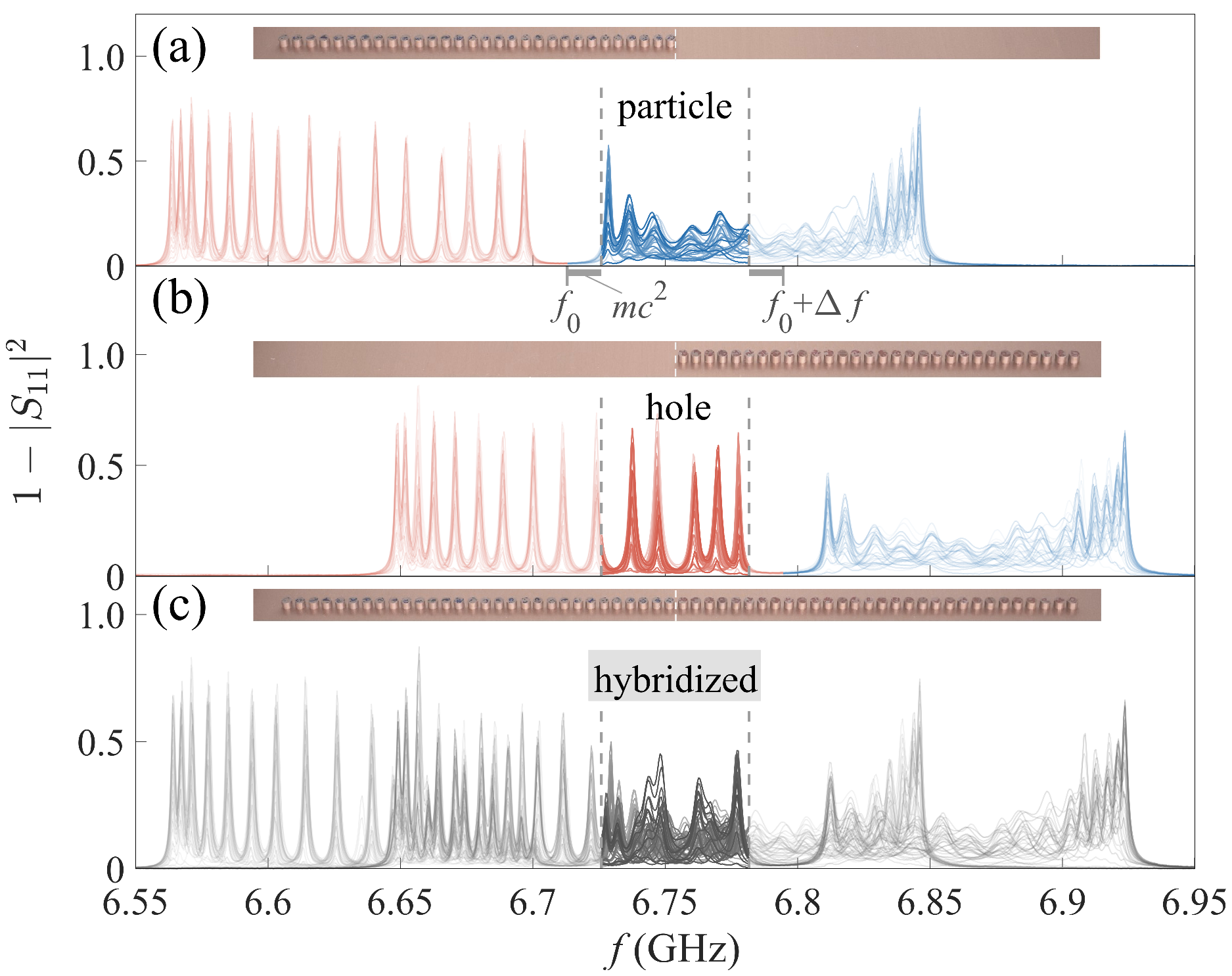}
     \caption{\label{fig:02} The reflection spectra for Experiment 1. The curves are measured on top of each resonator. (a) For the dimer chain of the $4$ mm resonators on the left side alone. (b) For the dimer chain of the $3.93$ mm resonators on the right side alone. (c) For the whole system. The region bounded by the dashed vertical lines indicates the Klein tunneling regime $(f_0 + mc^2) <f< (f_0 + \Delta f -mc^2)$. Blue (red) in (a,b) indicates resonances above (below) the gap, black in (c) indicates the hybridized resonances. Insets show the corresponding experimental dimer chains of dielectric resonators.  
     }
\end{figure}

We carry out a series of four experiments: Experiment 1 (E1): a symmetric case $(N_L, N_R) = (15, 15)$; E2: an independent realization of E1 with the same resonator order; E3: an independent realization of E1 but with a random permutation for both left and right resonators; E4: an asymmetric case (15, 9).

For each experiment, three sets of reflection spectra are measured for, e.g., the left $N_L$ dimers alone, the right $N_R$ dimers alone, and the whole system, as exemplified in Fig. \ref{fig:02} for E1. When focusing on the Klein tunneling regime, for the whole system [Fig. \ref{fig:02}(c)], more resonances occur, showing particle-hole hybridized states induced by the Klein tunneling.

The first two sets of the measured reflection spectra are to obtain the uncoupled particle and hole states close to the band gap, clearly seen in Fig. \ref{fig:02}(a,b), to extract the parameters. In particular, from the first set, the LDOS for the particle states above the band gap can be obtained, leading to the corresponding resonant frequencies and wavevectors $(f_n, k_{n})$. By fitting with $(f_n-f_0)^2=m^2c^4+c^2\hbar^2k_n^2$, the parameters $mc^2$, $c\hbar$ and $f_0$ are extracted. Similarly, from the second set, they can also be extracted by fitting with $(f-f_0-\Delta f)^2=m^2c^4 - \hbar^2c^2k^2$ for the hole states below the gap. Due to the change of radius, the spacing of the resonators on the right side has been fine tuned ($d_1 \rightarrow 9.96$ mm and $d_2 \rightarrow 10.48$ mm) to reach approximately the same parameter values as in the left side. From the four experiments, the extracted parameters are $mc^2 = 12.894$ MHz, $c\hbar /a_0 = 61.325$ MHz, and $f_0 = 6.713$ GHz, $V_0 = \Delta f = 81.5$ MHz (see Supplemental Material \cite{RefSM} for details).

Theoretically, with the obtained parameter values, we adapt the scattering matrix approach~\cite{Kottos1997, Sirko2012, Chen2021timedelay} to include the Klein tunneling effect for Dirac spinors. The unique feature is that the particle can propagate undamped in the barrier region, by turning into the hole. Thus instead of two channels in the scattering matrix for conventional tunneling~\cite{Bhullar2006}, there are four for the Hamiltonian $H$: 
\begin{equation*}
  e^{-{\mathrm i} k x} \binom{1}{-\xi}, e^{{\mathrm i} k x} \binom{1}{\xi},
  e^{{\mathrm i} \kappa (x-d)} \binom{-\zeta}{1}, e^{-{\mathrm i} \kappa (x-d)} \binom{\zeta}{1},
\end{equation*}
corresponding to the particles propagating to the left, right, and holes propagating to the left, right, respectively, 
where
$k= \sqrt{{E^2-m^2 c^4}}/{c \hbar}$ and $\kappa=\sqrt{{(V_0-E)^2-m^2 c^4}}/{c \hbar}$ are the wavenumbers for the particle and the hole; $\xi={c \hbar k}/{(E+m c^2)}$, $\zeta={c \hbar \kappa}/{(V_0-E+m c^2)}$. Consequently, the scattering matrix reads
\begin{equation}
   S = \begin{pmatrix}
   0  &  r_{pp} e^{{\mathrm i} ka}  & t_{ph} e^{{\mathrm i} ka}  &  0 \\
   r_L e^{{\mathrm i} ka}  &  0  &  0  & 0 \\
    0 & 0 & 0 &  r_R e^{-{\mathrm i}\kappa b} \\
    0 & t_{h p} e^{-{\mathrm i}\kappa b} & r_{hh} e^{-{\mathrm i}\kappa b} & 0
    \end{pmatrix},
\end{equation}
where $b=d-a$, $r_{pp}=r_{hh}=r={(\xi\zeta-1)}/{(\xi\zeta+1)}$ are the reflection and $t_{p h} = e^{{\mathrm i}\pi}t_{h p} = t =\sqrt{1-r^2}$ the transmission coefficients at the interface $x=a$, and the subindices specify the processes, e.g., ``${hp}$'' denotes transforming from particle to hole. $r_L =-e^{2{\mathrm i}\cdot \arctan{{\xi}}}$ and $r_R = - e^{-2{\mathrm i}\cdot \arctan  {{\zeta}}}$ are the reflection coefficients off the left and right boundaries, respectively. These transmission and reflection coefficients possess unusual phases ($\pi$, $2\arctan{{\xi}}$ and $-2\arctan  {{\zeta}}$) due to the spinor nature of the Dirac particles. The quantization condition is $\det (1 - S)=0$ \cite{Kottos1997, Bhullar2006}, leading to $\cos(\phi_a -\phi_b)=- r \cos(\phi_a + \phi_b)$, where $\phi_a = ka+\arctan{{\xi}}$, and $\phi_b = \kappa b+\arctan{{\zeta}}$. This yields the quantized levels $k_n$ (or $E_n$) and the corresponding eigenwavefunctions $\Psi_n$ of the system.

Figure \ref{fig:03} plots the LDOS (proportional to $|\psi|^2$) for E1 derived from the measured reflection spectrum in Fig. \ref{fig:02} only for the Klein tunneling regime. All the three cases follow well the boundary conditions, i.e., $\psi^{(B)}$ is close to zero at the left end, while $\psi^{(A)}$ is close to zero at the right end. The discrete eigenlevels from the theory are overlayed as the horizontal dashed lines, exhibiting an excellent agreement with the data. The observed Klein tunneling states (middle column) are clearly quantized.

There are 5 quantized particle (hole) states identified in the left (right) column. The particle (hole) manifests itself by increasing (decreasing), one by one, the number of peaks in the LDOS map when the frequency increases. The dependence of the quantum number on frequency originates from the quantization condition. For the whole system, due to Klein tunneling across the step potential, the 5 particle states and the 5 hole states are coupled and form 10 hybridized bound states. They inherit both the particle and hole's features. Consequently, when the wavenumber is small for the particle, it is usually large for the hole, and vice versa, in contrast to conventional tunneling induced hybrid orbitals \cite{Hund1927, Merzbacher2002}. Thus the quantized Klein tunneling states exhibit heterogeneous particle-hole wavefunctions.

\begin{figure}[t]
\includegraphics[width=\linewidth]{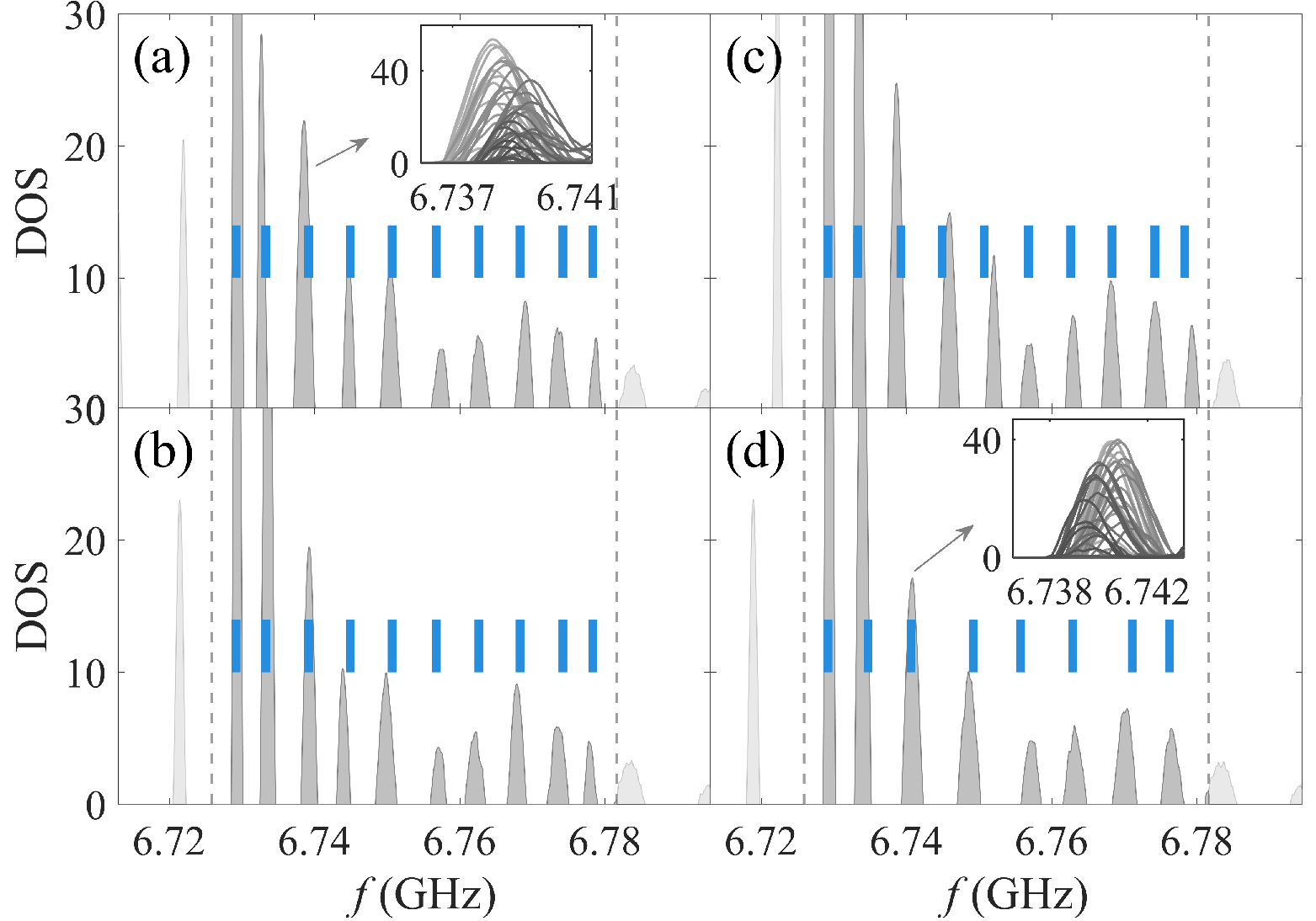}
\caption{\label{fig:04} Comparison between the measured DOS from the four experiments and the theoretical quantized levels ($f_0 + E_n$, vertical blue bars). (a-d) are for E1 to E4, respectively. The dashed vertical lines bound the Klein tunneling regime. Insets in (a) and (d) show the LDOS measured at all the resonators for representative quantized levels.
}
\end{figure}

\begin{figure}[t]
\includegraphics[width=\linewidth]{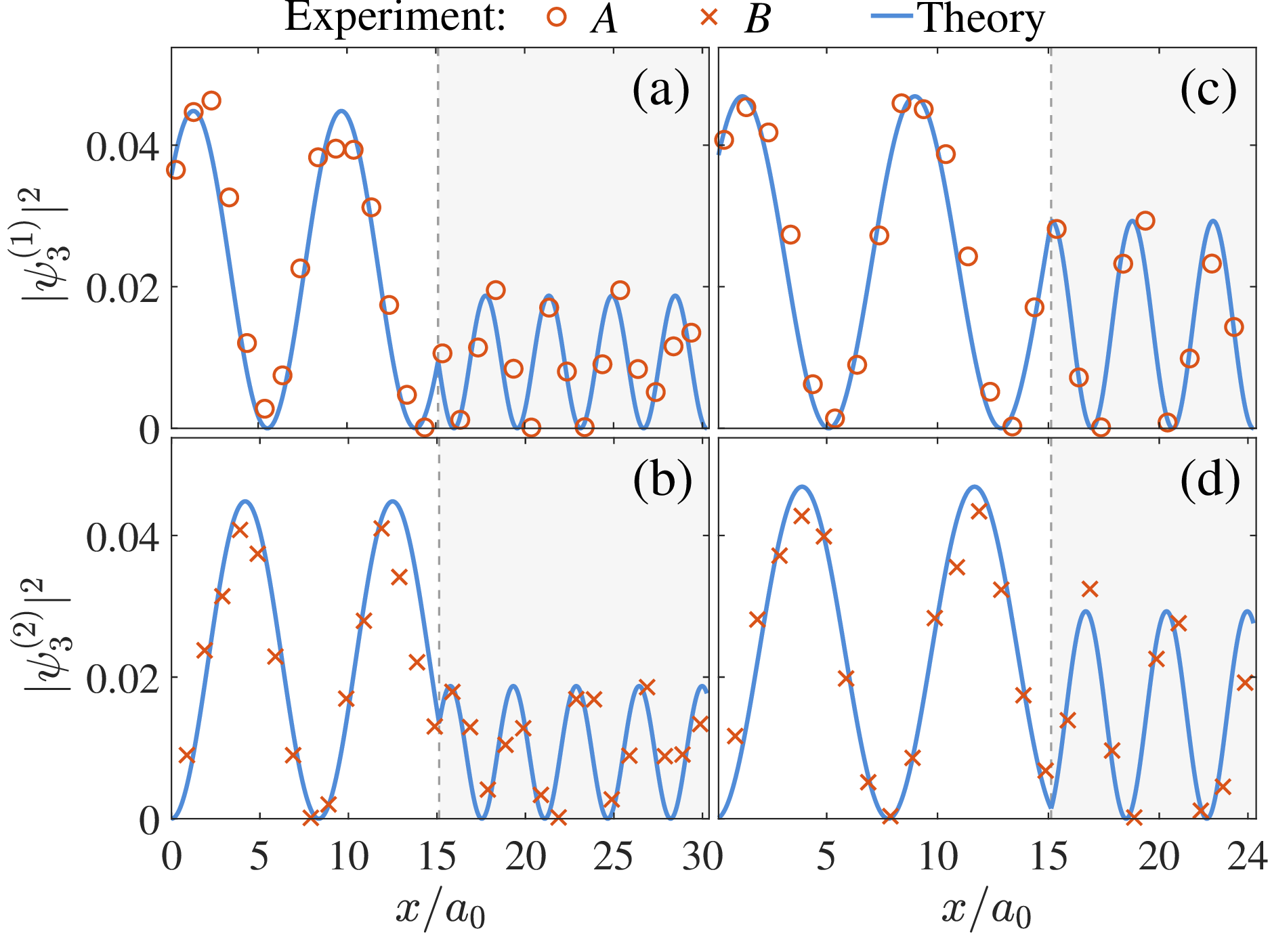}
\caption{\label{fig:05} Demonstration of the experimental and theoretical quantized spinor wavefunctions. (a,b) show the 3rd mode of E1, and (c,d) show the 3rd mode of E4. (a,c) display the wavefunction intensity for sublattice $A$ (the first component), (b,d) are for sublattice $B$ (the second component). The vertical dashed lines mark the position of the potential step.
}
\end{figure}

Figure \ref{fig:04} plots the measured DOS for the whole system from the four experiments, together with the theoretical predictions of quantized levels using the extracted parameters. Note that since the resonances for the hole states are so strong \cite{Bellec2013,Kuhl2010,Cl0Wavefront} that they are less responsive to the particle-hole binding, here the DOS is derived from the particle states only residing on the left domain of the whole system. One can see that the DOS peaks agree with the theory well for all the four experiments. The insets of Figs. \ref{fig:04}(a) and \ref{fig:04}(d) show the LDOS for two representative levels. The peaks are well separated for determining the spinor wavefunction intensities \cite{Bellec2013, Reisner2023, Bellec2013b}. The results are shown in Fig. \ref{fig:05}. For sublattice $A$, the theoretical prediction is $|\psi^{(1)}_3(x)|^2$, while for sublattice $B$, it is $|\psi^{(2)}_3(x)|^2$, where $\psi_3 = U \Psi_3$ is obtained from the Dirac equation. There is a great agreement between the experimental data and the theory. The data even follow the abrupt changes of the slope across the interface, visualizing the Klein tunneling wavefunction with an unprecedented level of accuracy. This evidences that the realized step potential is sufficiently abrupt, and also imply that the experiments capture, intriguingly, the unusual transmission and reflection phases due to the Dirac spinor wavefunctions correctly. Our experiments reveal, unambiguously, the quantized levels and states due to the Klein tunneling induced particle-hole hybridization.

{\it Conclusion.}---The direct observation of quantized levels and Klein tunneling states in microwave realizations of a bounded Dirac system has been demonstrated. The unique feature lies in the particle-hole hybridization via Klein-tunneling. By capitalizing on the great flexibility of the microwave setup, we have created the step potential for the Dirac equation of sufficient abruptness, and the hard-wall confinement as well. They are the key to reveal the unusual discrete Klein tunneling spectrum and states. Our results are crucial to the understanding of Klein tunneling in bounded Dirac systems. 

The amplitude and wavelength of the quantized Klein tunneling states are remarkably different from particle to hole across the potential step. 
The realized abrupt step potential can be modified readily to account for more complex step configurations. Thus our experiment offers a versatile platform for exploring relativistic particle-hole physics with the capability of tailoring heterogeneous spinor waveforms. These features carry over naturally to different frequency ranges of photonic crystals for potential applications.

\begin{acknowledgments}
This work was supported by National Key R\&D Program of China under Grant No. 2023YFA1407100, by NSFC under Grants No. 12105125, No. 12175090, No. 11775101, and No. 12247101, and by the 111 Project under Grant No. B20063.
\end{acknowledgments}

\bibliography{paper}

\begin{thebibliography}{54}
\bibitem [{ref()}]{reff0}%
  \BibitemOpen
  \href@noop {} {}\bibinfo {note} {In principle, if there are only the nearest
  neighbor coupling in the dimer chain, $f_0 = \nu_0$. When the next nearest
  neighbor coupling is non-negligible, $f_0$ will have a small shift, while the
  other relevant parameters are barely changed.}\BibitemShut {Stop}%
\end{thebibliography}

\end{document}